# The True-Twin microcalorimeter: a proof-of-concept experiment

Luca Oberto, Luciano Brunetti, and Marco Sellone

We present a proof-of-concept experiment to realize microwave primary power standard with a true-twin microcalorimeter. Double feeding line microcalorimeters are widely used by National Metrology Institutes. A drawback concerns the system calibration: traditional processes changes measurement conditions between system characterization and the measurement stage. Nevertheless, if the feeding lines are made twin, a measurement scheme that avoids separate characterization can be applied, equations simplify and time consumption is halved. Here we demonstrates the feasibility of the idea. The result of an effective efficiency spectroscopy of a thermoelectric power sensor is compared with figures obtained with well established methods.

*Introduction*: microcalorimeters were introduced in the late 1950s for the realization of the electromagnetic power standard at high frequency (HF) [1]. They are calorimetric systems adjusted for the measurement of small power ratios (typically about 1 mW) that allow the calibration of a root mean square (r.m.s.) power sensor tracing measurements to the International System of Units (SI). National Metrology Institutes (NMIs) make use of coaxial and waveguide microcalorimeters mainly based on the bolometric detection, but the indirect heating thermoelectric technology demonstrated to be a strong competitor mainly because sensors are not downward frequency limited and are much less sensitive to environmental temperature oscillations [2].

*The True-Twin microcalorimeter concept*: microcalorimetric technique based on thermoelectric detectors is, now, well established [3,4] and we refer specifically to it. The leading concept is that the HF power supplied to a proper sensor is not completely converted to a dc voltage output proportional to the input power, but a small fraction is dissipated into the sensor mount producing heat. Conversely, losses at low frequency (LF, normally 1 kHz) are negligible so that the ratio between the LF power $P_{LF}$ and a HF power $P_{HF}$ that produces the same sensor output $U$ (that is, the power effectively measured remains steady) gives the effective efficiency $\eta_e$ of the sensor under calibration:

$$\eta_e = \left.\frac{P_{LF}}{P_{HF}}\right|_{U=const}. \quad (1)$$

Because the temperature rise due to the HF losses is very small (of the order of some mK), a thermostat is needed in order to keep stable the measurement chamber temperature. The big challenge of this technique is the ability to distinguish between the HF losses produced into the sensor mount and the ones on the feeding line. This is done with a microcalorimeter calibration stage in which the feeding line is terminated by a load thermally equivalent to the power sensor under test, which an unitary effective efficiency can be attributed to. Unfortunately, the thermostat has to be opened to put in place the load, causing a change in the measurement conditions between the two steps of the sensor calibration procedure. This causes the worsening of the measurement uncertainty. Moreover, the system calibration stage doubles the time required by the whole process (the calibration of a sensor in 1 GHz steps, up to 26 GHz can require up to 2 month). In microcalorimeters fitted with a double feeding lines inset, a differential configuration is adopted in which an electrical thermometer senses the temperature difference between the sensor terminating one line and a thermally equivalent load terminating the other line. In the traditional measurement technique, only one line is energized with the HF and LF powers, whereas the other is used as a thermal reference. Nevertheless, if the two feeding lines are made twin both from the electromagnetic and the thermal point of view, we can realize a true-twin microcalorimeter. Both of them can be indifferently energized and this opens the way to a system self calibration; a reduction of the measurement uncertainty can be foreseen together with an improved realization of the primary HF power standards. Furthermore, because no previous calibration is needed, measurements become considerably less time consuming. To obtain the symmetry of the feeding lines, they must have the same length, mass, thermal conductance and electrical losses. Furthermore they have to be terminated with two thermally equivalent loads, one as device under test (DUT) and the other as dummy sensor (DS), respectively. If these conditions are met, the microcalorimeter can behave like an ideal and lossless one. In particular, if we use as DS a reflecting load, the microcalorimeter can be operated in a way that is independent on its own calibration coefficients. Consequently it is no longer necessary to open the measurement chamber for the hardware arrangements needed for their determination and the system thermal state is preserved. By supplying both the feeding lines with the HF and LF power alternatively, for the principle of superposition of the linear effects, a set of equations can be written that relates the power $P_S$ dissipated into the DUT or the DS and the line losses, $P_L$, to the asymptotic response $e$ of the electrical thermometer that measures the temperature at each load input. These equations turned out to be simplified due to the symmetry of the true-twin configuration that allows reducing the unknowns:

$$\begin{cases} e_{1-DUT} = \alpha R(K_A P_S + K_B P_L) \\ e_{1-DS} = \alpha R K_B P_L \\ e_{2-DUT} = \alpha R K_A P_S \\ e_{2-DS} = 0 \end{cases}, \quad (2)$$

in which $\alpha$ is the Seebeck coefficient of the two thermopiles, $R$ is a conversion constant, $K_A$ and $K_B$ are separating constants that describe how the losses divide between the feeding lines and the DUT or DS mount. The subscripts 1 and 2 refer to the presence of HF and LF power, respectively. Solving this system of equations in term of effective efficiency, as defined by Eq. (1), it can be seen that:

$$\eta_e = \frac{e_2}{e_1}. \quad (3)$$

It means that the DUT can be calibrated simply evaluating the ratio of the asymptotic voltage difference $e_1 = e_{1\text{-DUT}} - e_{1\text{-DS}}$ reached by the thermopile system when the DUT and the DS are supplied with the HF power and the corresponding asymptotic voltage difference $e_2 = e_{2\text{-DUT}} - e_{2\text{-DS}}$ obtained when both line loads are fed with the LF power. This relation is extremely simple if compared with the equations obtained operating the microcalorimeter asymmetrically with the traditional technique [3]. In facts, it is reduced to a voltage ratio as in the case of an ideal microcalorimeter and the calibration stage is no longer necessary.

*Proof-of-convept*: measurements have been performed from dc up to 26.5 GHz with the INRIM microcalorimeter [5] whose core is schematized in Fig 1. The two feeding lines are realized by using coaxial cables of the same material, length and dimensions together with specifically designed thermally insulating sections. Their electromagnetic behavior has been checked with transmission coefficient measurement. It resulted always greater than 0.9 and the relative difference between the two was always lower than 0.2 %. The DS is realized by short-circuiting the input connector of a sensor of the same type of the DUT. The measurement sequence starts, after thermal stabilization, cycling the HF power between the DUT channel and the DS channel for every measurement frequency having attention to halve the power when supplying the short-circuited DS in order to maintain constant the feeding line losses. After that, the HF power is substituted with an equivalent LF power according to (1) on both channels. The thermopile output recorded after about three time constants is shown in Fig. 2 in which the switching effect is evident. Applying Eq. (3), an effective efficiency spectroscopy can be obtained. Measurements have been performed in a wide frequency band where the used microcalorimeter characteristics and the DUT effective efficiency are well known [6]. A comparison of the DUT effective efficiency $\eta_{e\text{-ref}}$ determined through traditional measurements with the one evaluated with the true-twin microcalorimeter experiment $\eta_{e\text{-twin}}$ is presented in Tab. 1. Δ represents the relative difference between the two values. Methods agree at a degree of about 0.2 % up to 10 GHz. At higher frequencies the agreement worsens even if it is still good. This can be explained with an energetic unbalance of the feeding lines that is more evident when the HF losses are greater. Moreover, the true-twin theory assumes perfect impedance match of the DUT and unitary reflection coefficient of the DS. The real case is slightly different but it can be taken into account in more refined experiments.

*Conclusion*: this proof-of-concept experiment confirms the validity of the true-twin microcalorimeter principle. Difference observed between the true-twin results and traditional microcalorimetric measurements are interpreted as due, mainly, to residual thermal and electromagnetic imbalance of the feeding paths that can be minimized with *ad-hoc* optimization of the microcalorimeter structure. Other contributions arise from the impedance matching of the device under test and from imperfections of the short circuit applied to the dummy sensor. Anyway they can be taken into account in the mathematical models but seem to have less influence. With this method, lower uncertainty should be attained allowing a better realization of the HF primary standard. Moreover, the overall time consumption can be 50 % reduced because microcalorimeter calibration measurements are no longer necessary.

**Authors' affiliations:**
L. Oberto, L. Brunetti and M. Sellone (Electromagnetism Division, Istituto Nazionale di Ricerca Metrologica, strada delle Cacce 91, 10135, Torino, Italy)

E-mail: l.oberto@inrim.it

**Figures and table captions:**

Fig. 1   Schematic of the double feeding line microcalorimeter core with thermopile connections.

Fig. 2  Switching effect in the thermopile output voltage.

Tab. 1  Measurement results.

Figure 1

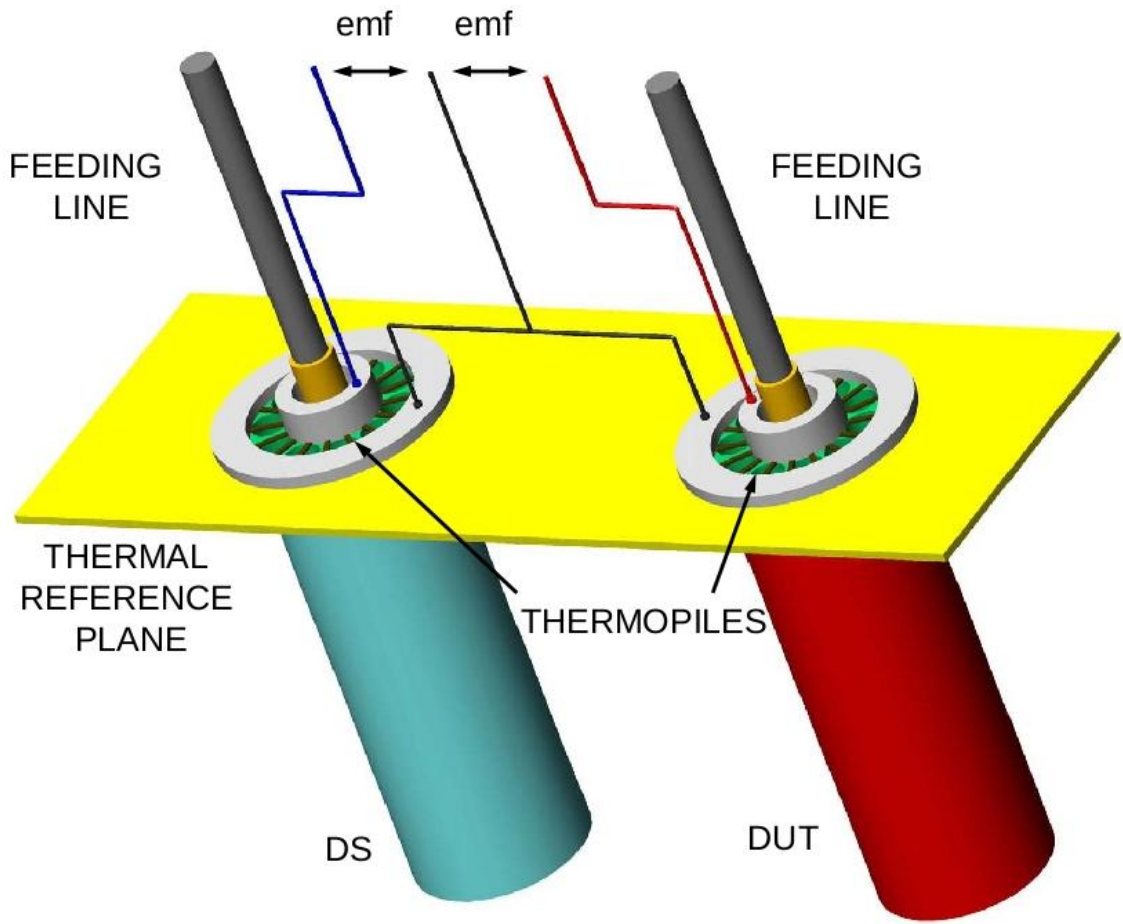

Figure 2

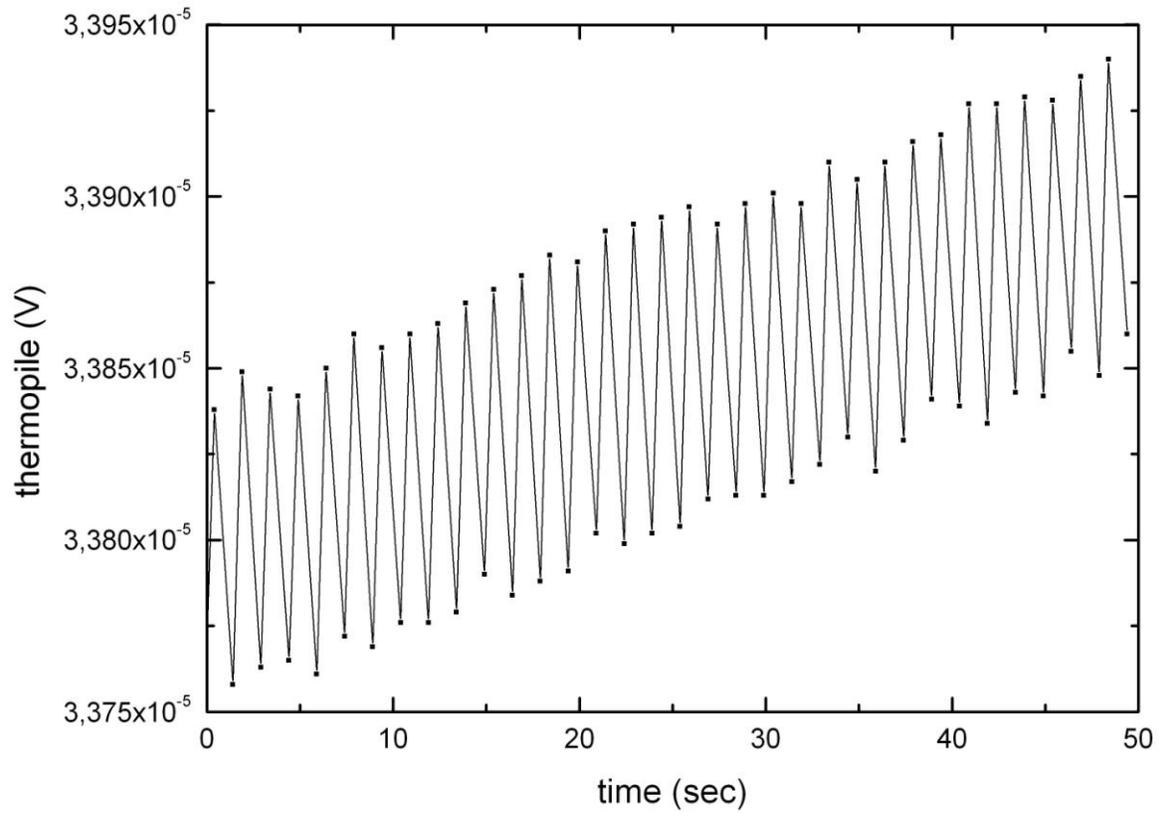

Table 1

| f / GHz | $\eta_e$ / % | $\eta_{e\text{-twin}}$ / % | Δ / % |
|---|---|---|---|
| 0.05 | 99.12 | 99.24 | 0.12 |
| 1 | 98.43 | 98.59 | 0.16 |
| 10 | 94.28 | 94.12 | -0.17 |
| 20 | 89.25 | 90.00 | 0.83 |
| 26 | 88.76 | 90.12 | 1.51 |